\documentclass{mem}
\usepackage{natbib}\usepackage{txfonts}\usepackage{balance}
\usepackage{graphicx}
\usepackage[a4paper]{hyperref}
\idline{79}{0}
\begin{document}
\def\teff{$T\rm_{eff }$}
\def\kms{$\mathrm {km s}^{-1}$}
\setcounter{page}{1201}

\title{
Galaxy activity influenced by the environment
in the cluster of galaxies Abell~85
}

   \subtitle{}

\author{
C.A.\,Caretta\inst{1},
J.M.\,Islas-Islas\inst{1},
J.P.\,Torres-Papaqui\inst{2},
R.\,Coziol\inst{1},
H.\,Bravo-Alfaro\inst{1}
\and H.\,Andernach\inst{1}
          }

  \offprints{C. Caretta}

\institute{
Departamento de Astronom\'ia --
Universidad de Guanajuato, 
Cjon. de Jalisco, s/n,
36000, Guanajuato (Guanajuato), M\'exico
\and
Coordinaci\'on de Astrof\'isica --
Instituto Nacional de Astronom\'ia \'Optica y Electr\'onica,
Luis Enrique Erro, 1,
72840, Tonantzintla (Puebla), M\'exico 
\\
\email{caretta@astro.ugto.mx}
}

\authorrunning{Caretta et al. }

\titlerunning{Galaxy Activity in Cluster}

\abstract{We analyse the relation between the dynamical state
of a cluster of galaxies and the activity (star formation 
and AGN) of its members. 
For the case of Abell~85 we find some evidence for an
enhanced activity of both types in substructures which are
in the early stage of merging with the cluster.

}
\maketitle{}

\section{Introduction}

It is widely accepted that the environment has strong 
influence on the formation and evolution of galaxies. 
One of the most established 
examples is the morphology-density relation \citep[e.g., ][]{Dre80}. 
The question then is: which are the mechanisms that cause this 
influence, and specifically, how do these mechanisms affect 
the activity of the galaxies -- probably one of the most important 
characteristics related to galaxy evolution? 
Here we present some results on a case study, the cluster 
Abell~85, of the influence of the environment on the star formation 
(SF) and AGN activity of its member galaxies.

\section{The Cluster Abell 85}

\begin{figure}[t!]
\resizebox{6.5cm}{!}{\includegraphics[clip=true]{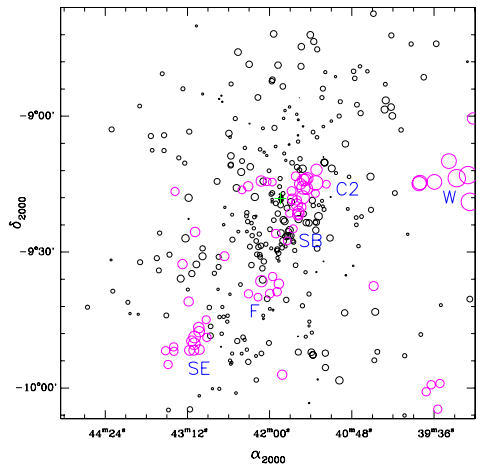}}
\caption{\footnotesize
Distribution of known member galaxies in A85 (367) with positions 
represented by circles of size proportional to $\delta$. 
Probable substructures, associated with concentrations of magenta 
circles, are labelled (SE, F, SB, C2 and W, see text). 
The position of the cD is marked with a green plus sign.}
\label{Fdelt}
\end{figure}

The cluster of galaxies A85 is a moderately rich (R = 1), relatively 
nearby (z = 0.055) cluster. It has been extensively studied over 
all the electromagnetic spectrum, especially in the radio, optical 
and X-rays.  
Many works have revealed a dynamically young system undergoing 
the merging of smaller clumps of galaxies. 
In Fig.~\ref{Fdelt} we show the results of the 
application of the $\Delta$-test \citep{DeS88}
to the 3D distribution of 367 member galaxies in A85 
[projected positions from 
SuperCOSMOS data \citep{Ham01} and radial velocities from an updated
version of the compilation by \citet{And05}]. 
This test searches for deviations of the local 
kinematics (based on the 10 nearest neighbors for each galaxy) and the 
global one (called $\delta$). We identify five substructures in this 
cluster based on the concentration of galaxies with deviant kinematics 
(displayed in magenta in Fig.~\ref{Fdelt}):

\noindent {\bf C2}: this substructure is composed of two clumps: the 
northern one, which is centered on the 2$^{nd}$-brightest galaxy; and the 
southern one, identified in a Chandra X-ray image \citep{Kem02};

\noindent {\bf SB} (South Blob): identified originally from X-ray 
imaging \citep{Kem02}; its kinematics is not particularly deviant 
but it represents the second highest density peak of the cluster;

\noindent {\bf F}: X-ray filament: also identified in X-rays \citep{Dur98};

\noindent {\bf SE}: this substructure is projected in the direction of the 
background cluster A87 (at z = 0.130);

\noindent {\bf W}: probably a clump that started interaction with A85 
most recently.

Details of this analysis are presented in (Bravo-Alfaro et al. 2008, 
in prep.).

\section{Classification  of Galaxies}

\begin{figure}[b!]
\resizebox{6.5cm}{!}{\includegraphics[clip=true]{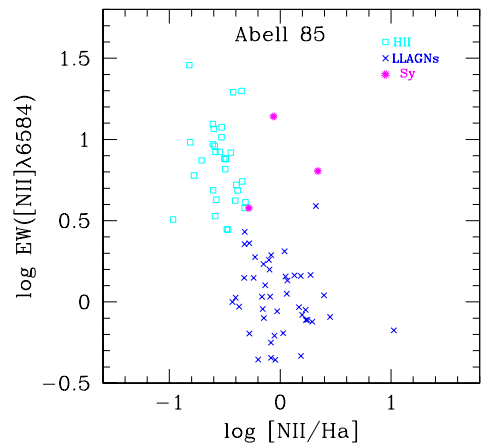}}
\caption{\footnotesize
Diagnostic diagram showing NII equivalent width vs the 
log([NII]/H$_\alpha$), which is not affected by absorption of the 
intragalactic medium and separates efficiently the AGN, LLAGN
(+ LINER) and HII populations (see Coziol et al. 1998).}
\label{dCoz}
\end{figure}

\begin{figure}[t!]
\resizebox{6.5cm}{!}{\includegraphics[clip=true]{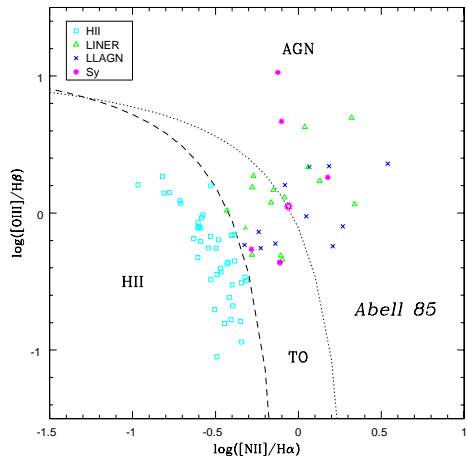}}
\caption{\footnotesize
Baldwin et al. (1981, BPT) diagram for A85 emission-line galaxies. 
The different types are color coded based on the previous diagram 
and visual inspection. Note that most LLAGNs do not present [OIII]
and/or H$_{\beta}$ with S/N $\ge 3$, and do not appear in this 
diagram.}
\label{BPT}
\end{figure}

For the analysis of activity (SF and AGN) we used an homogeneous 
sample of 232 spectra of A85 galaxy members (63\% of known members) 
available from SDSS \citep[e.g., ][]{Ade07}.
Using the \textsc{starlight} routine \citep{Cid05} we obtained the 
population synthesis of these spectra. Subtracting the 
respective synthetic stellar spectrum from each observed one we could 
fit and measure the emission lines. Only lines with S/N $\ge 3$
were used.
Two diagnostic diagrams \citep{Coz98,BPT81} 
and an exhautive visual inspection 
were used in order to 
separate the emission line galaxies into HII and AGN domains 
(Figs. \ref{dCoz} and \ref{BPT}). 
Our results show that 17\% of A85 members are HII, 17\% are
LLAGNs, 5\% are LINERs and 2.6\% are high-luminosity AGNs,
the remaining being galaxies without emission-lines 
(noEL).

This AGN fraction in a cluster is much higher than previously 
reported, especially before the era of large spectroscopic 
surveys, when the estimates showed a decrease of the number of AGNs in 
clusters with respect to the field \citep[e.g., ][]{Gis78}. 
A recent X-ray survey of A85 \citep{Siv08} 
detected 4 high luminosity AGNs in a sample of 170 member galaxies 
of A85 (2.4\%), in accordance with our results.

%

\section{Environmental Effects} 

\begin{figure*}[t!]
\begin{center}
\resizebox{8cm}{!}{\includegraphics[clip=true]{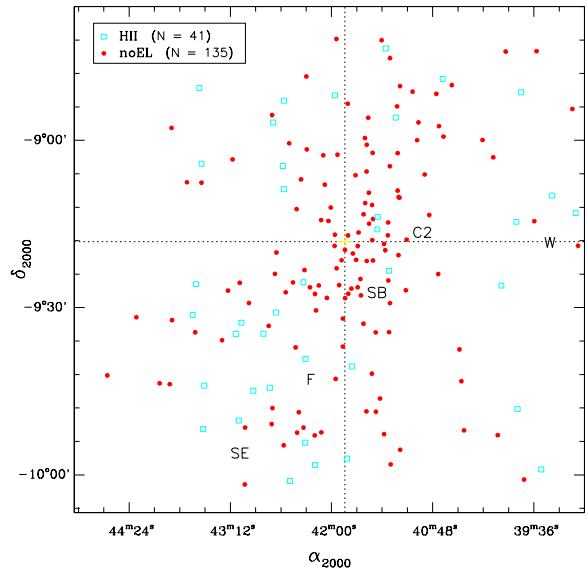}}
\caption{\footnotesize
Distribution of noEL and HII galaxies in our sample of A85 members
with spectra.
The substructures are coded as in Fig.~\ref{Fdelt}. 
The quadrants are marked with respect to the cD.}
\label{Fclas2}
\end{center}
\end{figure*}
\begin{figure*}[b!]
\begin{center}
\resizebox{8cm}{!}{\includegraphics[clip=true]{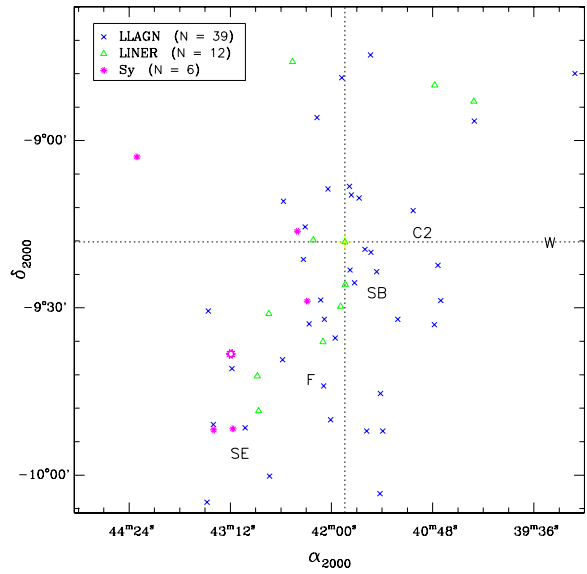}}
\caption{\footnotesize
Distribution of AGNs in our sample of A85 galaxies. Codes 
are the same as in previous figures. The larger magenta symbol
represents the only Sy1 galaxy in our sample.}
\label{Fclas3}
\end{center}
\end{figure*}

We correlate the distribution of 
our sample of galaxies spectroscopically classified with the 
substructures in A85.
In Fig. \ref{Fclas2} one can see that the noEL
present a fairly homogeneous distribution, concentrated toward 
the center as expected. On the other hand, the HII galaxies show some 
concentration in the SE quadrant. 
This is the region comprising the majority of the substructures 
found in A85, namely the SE, the X-ray filament and the SB.
Since this ridge of clumps is poblably the most dynamically active 
part of the cluster in recent times, representing a sequence of 
future merging structures to the main body of the cluster, we are 
lead to conclude that there is some stimulation of SF
activity by this dynamical status.
More than that, the AGN activity (Fig. \ref{Fclas3}), both the 
low energy phenomenon (LLAGNs and LINERs) and high energy one (Sy2), 
present a more enhanced concentration around the ridge. 
All of the Sy2 galaxies are located on the eastern side of 
A85, and two of them are members of the SE substructure.
This relative overdensity of active galaxies inside the substructure SE 
suggests that activity is favored inside substructures that are 
beginning to interact with the cluster. 
We have also started to study the complete SF history of 
our sample galaxies using \textsc{starlight}, which will be
presented elsewhere. 

\section{Conclusions}

The main results of this work are:
(a) we found a very large number of LLAGNs \citep{Ho93} 
in A85, with a similar abundance to the one found in compact groups of 
galaxies \citep{Coz98}; this suggests the LLAGN phenomenon is very 
common in dense environments;
(b) there is some evidence of enhanced activity,
both SF and AGN, in substructures in the 
early stage on their process of merging with the cluster;
(c) we suggest that the distribution of active galaxies 
may be used to search for the presence of subtructures in clusters of 
galaxies.

\begin{acknowledgements}
We are grateful to CONACyT and Universidad de Guanajuato for supporting 
this project. We also acknowledge the use of SDSS data (the SDSS Web 
Site is http://www.sdss.org/).
\end{acknowledgements}

\bibliographystyle{aa}

\begin{thebibliography}{}

\bibitem[{Adelman-McCarthy et al. (2007)}]{Ade07} 
Adelman-McCarthy, J.K., et al. 2007, ApJS 172, 634.

\bibitem[{Andernach et al. (2005)}]{And05} 
Andernach, H., et al.
2005, ASPC 329, 283.

\bibitem[{Baldwin, Phillips \& Terlevich (1981)}]{BPT81} 
Baldwin, J.A., Phillips, M.M., Terlevich, R. 1981, PASP 93, 5.


\bibitem[{Cid Fernandes et al. (2005)}]{Cid05} 
Cid Fernandes, R., et al.
2005, MNRAS 358, 363.

\bibitem[{Coziol et al. (1998)}]{Coz98} 
Coziol, R., et al.
1998, ApJ 493, 563.

\bibitem[{Dressler (1980)}]{Dre80} 
Dressler, A. 1980, ApJ 236, 351.

\bibitem[{Dressler \& Shectman (1988)}]{DeS88} 
Dressler, A., Shectman, S.A. 1988, AJ 95, 985.

\bibitem[{Durret et al. (1998)}]{Dur98} 
Durret, F., et al.
1998, A\&A 335, 41.

\bibitem[{Gisler (1978)}]{Gis78} 
Gisler, G.R. 1978, MNRAS 183, 633.

\bibitem[{Hambly et al. (2001)}]{Ham01} 
Hambly, N.C., et al. 2001, MNRAS 326, 1279.

\bibitem[{Ho et al. (1993)}]{Ho93} 
Ho, L.C., Shields, J.C., Filippenko, A.V. 1993, ApJ 410, 567.

\bibitem[{Kempner et al. (2002)}]{Kem02} 
Kempner, J.C., Sarazin, C.L., Ricker, P.M. 2002, ApJ 579, 236.

\bibitem[{Sivakoff et al. (2008)}]{Siv08} 
Sivakoff, G.R., et al. 2008,
arXiv0804.3797.

\end{thebibliography}

\end{document}